\documentclass[12pt]{article}

\usepackage{namedcaptions}  

\usepackage[xcharter]{newtxmath}

\usepackage{setspace}
\usepackage{booktabs}
\usepackage{rotating}  
\usepackage{xcolor}
\usepackage[normalem]{ulem}
\providecommand{\explain}[1]{\medskip\par{\textbf{Explanations:} #1}}
\newcommand{\xpanelpre}{0\baselineskip}
\newcommand{\xpanelpost}{0\baselineskip}
\newcommand{\panel}[2]{\vspace{\xpanelpre}{\medskip\centering\textbf{Panel~#1:} #2}\vspace{\xpanelpost}}
\usepackage[labelfont=bf,position=top]{caption}
\usepackage{array}  
\def\GobbleColumnStart{\setbox0\hbox\bgroup}
\let\GobbleColumnStop\egroup
\newcolumntype{n}{@{}>{\GobbleColumnStart}c<{\GobbleColumnStop}}    
\def\GobbleColumnStart{\setbox0\hbox\bgroup}
\let\GobbleColumnStop\egroup
\newcolumntype{i}{@{}>{\GobbleColumnStart}c<{\GobbleColumnStop}}    
\newcolumntype{d}{>{\centerdots}c<{\endcenterdots}}
\newcolumntype{S}{!{\qquad}}
\newcolumntype{s}{!{$\;\;$}}
\def\testnext#1#2{#1\ifx-#2\color{red}-\else\ifcat#20\color{blue}\else\fi\fi#2}
\newcolumntype{R}{>{\testnext}r<{}}
\newcolumntype{D}{>{\scriptsize\testnext}r<{}}
\def\testnextopenparen#1#2{#1\ifx-#2\color{red}(-\else\ifcat#20\color{blue}\else\fi\fi#2}

\makeatletter
\newenvironment{basectabular}[1]{%
  \skip@=\baselineskip
  #1%
  \baselineskip=\skip@
  \vspace{0.5\baselineskip}
  \ctabular
}{\endctabular\ignorespacesafterend\vspace{0.75\baselineskip}}
\makeatother
\newenvironment{ctabular}{\begin{center}\begin{tabular}}{\end{tabular}\end{center}}
\newenvironment{smctabular}{\begin{basectabular}{\small}}{\end{basectabular}}

\newcommand{\mc}{\multicolumn}

\usepackage{hyperref}
\makeatletter
\def\@urlcolor{darkblue}
\def\@linkcolor{darkred}
\hypersetup{urlcolor=darkblue}
\hypersetup{linkcolor=darkred}
\makeatother
\usepackage{amsmath}
\usepackage{needspace}

\definecolor{verylightgray}{rgb}{0.92,0.92,0.92}
\definecolor{veryverylightgray}{rgb}{0.96,0.96,0.98}

\usepackage[backend=biber,style=authoryear,uniquename=false,autopunct=true,language=USenglish]{biblatex}
\bibliography{kossin}

\DeclareDocumentCommand{\ig}{m}{\includegraphics[width=0.35\textwidth]{figs/#1}}
\DeclareDocumentCommand{\igb}{m}{\includegraphics[width=0.43\textwidth]{figs/#1}}

\DeclareDocumentCommand{\sixfig}{m}{
  \footnotesize
  \begin{ctabular}{c@{}c@{}c}
    \uline{North Atlantic (NA)}  & \uline{Southern Indian (SI)} & \uline{South Pacific (SP)}   \\[1ex]
    \ig{#1-1NA-crop}              &    \ig{#1-2SI-crop}           & \ig{#1-3SP-crop}              \\[2em]
    \uline{Eastern Pacific (EP)} & \uline{North-W Pacific (NP)} & \uline{Northern Indian (NI)} \\[1ex]
    \ig{#1-4EP-crop}              & \ig{#1-5WP-crop}   & \ig{#1-6NI-crop}              \\[3em]
    &\uline{Sum-Total All Basins} \\[1ex]
    & \ig{#1-0ALL-crop} \\
  \end{ctabular}
}

\DeclareDocumentCommand{\fourfig}{m}{
  \footnotesize
  \begin{ctabular}{c ss c}
    \uline{North Atlantic} & \uline{Southern Indian} \\
    \igb{#1-1NA-crop}       & \igb{#1-2SI-crop}        \\[2em]
    \uline{South Pacific}   & \uline{North-W Pacific} \\
    \igb{#1-3SP-crop}        &\igb{#1-4EP-crop}         \\[2em]
    \mc2c{\uline{Sum-Total All Basins}} \\[1ex]
    \mc2c{\igb{#1-0ALL-crop}} \\
  \end{ctabular}
}

\begin{document}

\title{Disaggregating the Increase in Tropical Cyclone Intensity in Kossin et al. (PNAS 2020)}

\author{Ivo Welch}

\maketitle

\begin{abstract}
  \textcite{kossin2020global} successfully test (over the last four decades) the prediction of climate-change models that conditional tropical cyclone intensity (the frequency of major [category~3-5] cyclones divided by the frequency of all cyclones) should increase.  The highest relative proportional increase in incidence occurred in category~3 cyclones.  The highest relative proportional decrease occurred in category~5 cyclones.  This inference is not due to different treatment of measurement error.  It is plausibly due to lower power for relatively rarer category~5 cyclones.
\end{abstract}

It is well known that the number of total cyclones has decreased in recent decades.   This is likely due to decadal variations and not evidence against climate-change models.  Many global-climate models have been predicting not an increase in the number of cyclones, but an increase in the intensity of cyclones.  In an important recent paper, \textcite{kossin2020global} successfully tests this prediction.  They state that ``theory and numerical models consistently link increasing TC intensity to a warming world, but confidence in this link is compromised by difficulties in detecting significant intensity trends in observations. These difficulties are largely caused by known heterogeneities in the past instrumental records of TCs. Here we address and reduce these heterogeneities and identify significant global trends in TC intensity over the past four decades. The results should serve to increase confidence in projections of increased TC intensity under continued warming.''

My note disaggregates this analysis into different category-strength hurricanes.  It confirms Kossin, but also shows that the intensity increase was not monotonic in category.  Under additional assumptions about measurement error, further results obtain.

\section{Replication}

The ADT-HURSAT record created by and examined in \textcite{kossin2020global} was transformed to be homogenous in time and basin.  Their paper mitigates other uncertainties by focusing only on estimates of major (C3-C5) and minor (C1-C2) cyclones, where ``C'' means category).  This is also appropriate for their emphasis on changes in the stronger tropical cyclones.  They describe trends in six basins: The North-Atlantic (NA), the Eastern-North Pacific (EP), the Western-North Pacific (WP), the Northern-Indian (NI), the Southern-Indian (SI) and South-Pacific (SP).  My paper uses their state-of-the-art data and thus follows their choices.

\instbl{tbl:descriptive}

The descriptive statistics for the data set in \textcite{kossin2020global} are in Table~\ref{tbl:descriptive}.  In an average three-year period, the mean number of cyclones per basin was 237$\pm$172.  The EP and WP were the two most active basins, the NI basin was the least active.  Cyclone numbers in the first third of the sample were larger than in the second.  Cyclones of 

\insfig{fig:b}

Figure~\ref{fig:b} shows the replication of \textcite[Figure 3]{kossin2020global} using their posted data, with only minor alterations.  Using ``C'' henceforth as abbreviation for Saffir-Simpson category, each point is the number of C3-C5 cyclones divided by the number of C1-C5 cyclones within a given three-year period.\footnote{The 1980 observation contains only 1979 and 1981 due to lack of satellite coverage.  Periods are named by the middle year.}  The plots are ordered by the magnitude of an OLS regression line slope with year as the dependent variable.  The bottom panel first aggregates all cyclones over all basins, and repeats the analysis.  Such an ``all'' index places more weight on basins with more observations (e.g., here ten times as much weight on WP as NI).

All plots show positive trends in this regression line.  \textcite{kossin2020global} use Theil-Sen lines, based on medians.  This only makes a difference for the Northern Indian basin, where OLS suggests a positive slope while Theil-Sen suggests a negative slope.  The Theil-Sen line is in line with the cooling Indian Ocean multidecadal variability, while the OLS line is not.

\section{Time-Series Changes By Cyclone Category}

\subsection{Measurement Error}

\renewcommand{\b}[1]{\beta_{#1 \leftarrow yr_t}}

\textcite{kossin2020global} describe ratios as being more reliable than absolute numbers.  This is the case, e.g., if measurement errors $\epsilon_{it}$ are identical in time for each basin and thus cancels,
\[ \frac{\epsilon_{it}\cdot m_{it}}{\epsilon_{it}\cdot m_{it} + \epsilon_{it}\cdot n_{it}} =
  \frac{m_{it}}{m_{it} + n_{it}} \quad \Rightarrow \quad \b{r_{it}} = \frac{cov(\,m_{it}+n_{it}, yr_{it}\,)}{var(\,yr_{it}\,)} \quad (\perp \epsilon_{it}) \;,\]
where $m_{it}$ are the number of major cyclones in basin $i$ at time $t$, $n_{it}$ are the number of minor cyclones, and  $\b{m}$ is the OLS slope regression of~$m$ cyclones on a year index~$yr$.  The slope of the log number of cyclones on the year index is still influenced by the $\epsilon$ measurement errors,
\[ \b{\log(\epsilon_{it} m_{it})} = \frac{cov(\,\log(\epsilon_{it} m_{it}), yr_{it}\,)}{var(\,yr_{it}\,)} = \frac{cov(\,\log(\epsilon_{it}), yr_{it}\,)}{var(\,yr_{it}\,)} + \frac{cov(\,\log(m_{it}), yr_{it}\,)}{var(\,yr_{it}\,)} \;.\]%
\[ = \b{\log{\epsilon_{it}}} \;+\; \b{\log{m_{it}}} \]
Thus, although reductions over time in the variance of the log measurement error do not bias the coefficient, time trends in log measurement errors can indeed bias this slope.  Specifically, if measurement improvements have increased the estimated number of observed major cyclones more than the estimated number of observed minor cyclones, it would manifest itself in Type-1 error (false acceptance).  If measurement improvement have decreased the estimated number of observed major cyclones more than the estimated number of observed minor cyclones, it would manifest itself in Type-2 error (false rejection).  I have no subject expertise to opine on time-series trends in observed cyclones of different types due to measurement changes.

However, just like the Kossin ratio metric, regardless of any relative time-trends induced by error, the difference between two slopes is not sensitive to $\epsilon$ measurement error:
\[ \b{\log(\epsilon_{it} m_{it})} - \b{\log(\epsilon_{it} n_{it})} = 
\frac{cov(\,\log(m_{it}), yr_{it})}{var(yr_{it})} - \frac{cov(\,\log(n_{it}), yr_{it})}{var(yr_{it})} \;.\]
\[ = \b{\log(m_{it})} \;-\; \b{\log(n_{it})} \qquad (\perp \epsilon_{it}) \]

Unfortunately, in cases in which there were no cyclones in a three-year period, the log change cannot be computed.  This is the case in many basins for C5 (and sometimes also C4) cyclones.

\subsection{Slope Estimates}

The decomposition into two components allows us to examine whether the ratio increase was due more to a decline in the number of minor cyclones or a rise in the number of major cyclones, or both.

\instbl{tbl:cslopes}

Table~\ref{tbl:cslopes} tabulates the OLS time slopes of the log-number of cyclones on the year index.  The two panels differ in their treatment of the first three-year observation, in which 1980 was missing.   In Panel~A, incidences are multiplied by 3/2 to account for the missing year of 1980.  In Panel~B, the first three-year observation is excluded altogether.

The left-most cell in both panels' first rows shows that the number of all categorized cyclones worldwide declined over these four decades.  As pointed out earlier, climate-change models did not predict an increase.  Thus, this is not evidence for or against an influence of warming on cyclone activity.

The log has transformed changes into relative proportional changes.  The strongest proportional worldwide increase occurred in C3 cyclones.  The strongest proportional worldwide decrease occurred in C5 cyclones.  The WP basin, which had the largest number of C5 cyclones, showed an even stronger decline.\footnote{Useful log numbers can only be computed for the WP and for the overall world, because other individual basins had cases in which the number of C5 cyclones in a given three-year period was zero.} Panel~B also shows that the absolute increase in C3 cyclones is sensitive to the inclusion of the first three-year period.  If there were no trends in measurement errors that systematically reduced the log number of cyclones counted, then the increase in intensity reported by Kossin et al. is entirely due to a reduction in the frequency of low-intensity cyclones.

The slope differences across different categories is largely unchanged by the first three-year treatment.

This raises the question why C5 cyclones did not show stronger proportional increases than C4 cyclones and why C4 cyclones did not show more frequent increases than C3 increases.  One possibility is that the climate models may predict that C3-C5 cyclone frequency would increase, but not necessarily C4-C5 frequency.  A more likely possibility is that this may well have been due to random sample noise --- four decades may not be sufficient to observe the expected relative increase in C5 cyclones.  Future years will clarify the picture.

The remaining rows confirm Kossin's within-basin evidence.  All basins show larger proportional increases in C345 cyclones than in C12 cyclones.  (Not shown, their ratio-increase inference is also quite robust with respect to different choices of cutoffs or years.)  However, the world-wide slope is larger when basins are equal-weighted than when they are weighted by the number of cyclones.  This is because the NA and SI basins, which show the strongest increases, contained only 27\% of the sample, while the EP and WP basins, which show the weakest increases, contained 60\% of the sample.  Nevertheless, given that all slopes are positive, the all-basin aggregate inference in Kossin et al. is easily confirmed.

Table~\ref{tbl:statsig} shows statistical tests of the differences in the first panel.  C1 cyclones increased significantly more than C3 and C4 cyclones (T of 3.49 and 2.92); and C3 cyclones increased significantly more than C2 cyclones.  The inference about the unexpected reduction C5 cyclones leaving in the first period just misses statistical significance, with $T= -1.98$ relative to C3 cyclones, $T= -1.76$ relative to C4 cyclones, and $T= -1.92$ relative to C3 and C4 cyclones.  It is insignificant if the first period is omitted.

\subsection{Log Number of Cyclones}

\insfig{fig:n}

Figure~\ref{fig:n} adds the log number of C3-C5 cyclones and the log number of C1-C2 cyclones to their equivalents from Figure~\ref{fig:b}.

Visual analysis suggests that C3-5 cyclones increased strongly from 1983 to 2005 in the NA basin.

The other basins display (mildly) more puzzling trends.  The SI basin experienced decreases in all cyclone activity since 1990 --- suggesting that despite its clear increase in cyclone intensity, the effects may have been less of a concern to local policy makers.  Cyclone activity also decreased in the SP and WP basins from about 1980 to about 2008.  However, both basins then saw strong increases in the 2010s, especially in major C3-C5 cyclones.

\section{Conclusion}

My note has analyzed some of the sources in trends in \textcite{kossin2020global}.  On a worldwide basis, the increase in the intensity of cyclones was not monotonic by category.  C5 cyclones did not increase relative to C4 cyclones which did not increase relative to C3 cyclones.  However, C3 cyclones increased greatly relative to C2 cyclones which increased greatly relative to C1 cyclones.  Under the additional assumptions of no systematic downward trend in measurement error, the increased intensity in the Kossin data set is due about equally to a reduction in the frequency of low-intensity cyclones and an increase in the frquency of high-intensity cyclones if the first three-year period is included; and due entirely to a reduction in the frequency of low-intensity cyclones if the first period is omitted.

Even though the number of C5 category has declined over these decades, to the extent that the world-wide decline in cyclone numbers was temporary and the increase in intensity will be persistent, policy-makers must not become complacent


\printbibliography


\clearpage

\begin{table}
  \newenvironment{mysub}{
    \begin{tabular}{csRs *{5}{R}}
      \toprule
           & All   & C1  & C2  & C3  & C4  & C5 \\
      \midrule}{\bottomrule\end{tabular}}

  \tblcaption{tbl:descriptive}{Descriptive Statistics by Basin and Category}

  \small

  \panel{A}{Sample Mean and Standard Deviations}

  \begin{ctabular}{cssc}
    Mean & Standard Deviation \\[1ex]
    \begin{mysub}
      ALL & 1,423 & 661 & 257 & 253 & 216 & 36 \\
    \midrule
      NA  & 184   & 88  & 34  & 32  & 27  & 3  \\
      SI  & 205   & 110 & 32  & 37  & 26  & 1  \\
      SP  & 148   & 73  & 27  & 26  & 19  & 3  \\
      EP  & 352   & 184 & 56  & 57  & 49  & 6  \\
      WP & 497   & 180 & 106 & 95  & 92  & 23 \\
      NI  & 37    & 27  & 3   & 5   & 2   & 0  \\
    \end{mysub}
  & 
  \begin{mysub}
    ALL & 233 & 106 & 56 & 56 & 44 & 14 \\
    \midrule
    NA  & 95  & 37  & 17 & 23 & 27 & 4  \\
    SI  & 77  & 42  & 13 & 21 & 12 & 1  \\
    SP  & 60  & 31  & 14 & 13 & 8  & 3  \\
    EP  & 136 & 66  & 27 & 22 & 27 & 4  \\
    WP & 95  & 45  & 23 & 18 & 22 & 15 \\
    NI  & 16  & 10  & 3  & 4  & 3  & 0  \\
  \end{mysub}
\end{ctabular}

\explain{This table shows means and standard deviations over thirteen 3-year periods.  For example, there were about 1,423 cyclones per three-year period in the Kossin sample, i.e., about 470 per year.  The average yearly season in the NA (North-Atlantic) brought about 60 cyclones in total.  The worldwide standard deviation was about 233 cyclones per three-year period.}

\end{table}

\begin{figure}[h]

  \figcaption{fig:b}{\textcite[Figure 3]{kossin2020global} Replicated}

  \sixfig{b}

\end{figure}

  
  \newenvironment{mytab}{
    \begin{smctabular}{c sRs s|s RRRRR s|s nnnnnnn nnR ss}
      \toprule
      \cmidrule(lr){3-7}
      \cmidrule(lr){8-12}
      \cmidrule(lr){13-17}
      & All   & C1    & C2    & C3    & C4 & C5 & C12 & C123 & C23 & C345 & C45 & C05 & C45-C1 & C45-C12 & C45-C123 & \makebox[1em]{C345 -- C12} \\
      \midrule
    }{\bottomrule
    \end{smctabular}}

\begin{table}[h]

  \tblcaption{tbl:cslopes}{Coefficients of Logged Number of Cyclones On Year}

  \panel{A}{Include First Two-Year Group (multiplied by 3/2), in Percent}

  \begin{mytab}
    0ALL & -0.27 & -0.81 & -0.09 & 0.67  & 0.27  & -1.23 & -0.61 & -0.34 & 0.28  & 0.35  & 0.05  & -0.42 & 0.85  & 0.65  & 0.39  & 0.95 \\
    \midrule
    1NA  & 2.64  & 1.35  & 3.26  & 4.61  & NA    & NA    & 1.77  & 2.29  & 3.90  & 4.80  & NA    & 1.08  & NA    & NA    & NA    & 3.04 \\
    2SI  & 0.09  & -0.82 & -0.35 & 2.27  & 1.97  & NA    & -0.75 & -0.28 & 0.80  & 2.20  & 2.12  & -0.07 & 2.94  & 2.87  & 2.40  & 2.95 \\
    3SP  & -1.98 & -2.49 & -2.14 & -1.37 & -1.34 & NA    & -2.34 & -2.12 & -1.75 & -1.31 & -1.17 & -1.55 & 1.32  & 1.17  & 0.95  & 1.03 \\
    4EP  & -0.70 & -1.15 & -0.45 & 0.01  & 0.14  & NA    & -1.01 & -0.83 & -0.20 & 0.08  & 0.12  & 0.77  & 1.27  & 1.13  & 0.95  & 1.08 \\
    5WP  & -0.48 & -0.71 & -0.27 & 0.33  & -0.72 & -2.39 & -0.55 & -0.32 & 0.03  & -0.38 & -1.02 & -1.38 & -0.31 & -0.47 & -0.70 & 0.17 \\
    6NI  & -0.63 & -0.97 & NA    & NA    & NA    & NA    & -0.78 & -0.69 & NA    & NA    & NA    & 1.14  & NA    & NA    & NA    & NA   \\
  \end{mytab}

  \panel{B}{Exclude First Two-Year Group, in Percent}

  \begin{mytab}
    0ALL & -0.65 & -1.11 & -0.53 & -0.03 & -0.06 & -1.27 & -0.95 & -0.74 & -0.27 & -0.12 & -0.24 & -0.68 & 0.87  & 0.71  & 0.50  & 0.82 \\
    \midrule
    1NA  & 3.15  & 2.00  & 3.41  & 4.75  & NA    & NA    & 2.27  & 2.77  & 4.04  & 5.30  & NA    & 1.67  & NA    & NA    & NA    & 3.03 \\
    2SI  & -0.78 & -1.68 & -0.84 & 0.73  & 1.17  & NA    & -1.54 & -1.17 & -0.08 & 0.96  & 1.32  & -0.90 & 2.99  & 2.86  & 2.48  & 2.50 \\
    3SP  & -2.35 & -2.85 & -2.67 & -2.13 & -1.08 & NA    & -2.75 & -2.58 & -2.39 & -1.60 & -1.00 & -1.67 & 1.85  & 1.75  & 1.58  & 1.15 \\
    4EP  & -1.48 & -1.64 & -1.59 & -1.18 & -1.15 & NA    & -1.61 & -1.53 & -1.36 & -1.19 & -1.21 & -0.03 & 0.42  & 0.40  & 0.31  & 0.42 \\
    5WP  & -0.67 & -0.89 & -0.44 & -0.08 & -0.88 & -2.00 & -0.73 & -0.55 & -0.24 & -0.57 & -1.08 & -1.55 & -0.19 & -0.35 & -0.53 & 0.16 \\
    6NI  & 0.10  & -0.41 & NA    & NA    & NA    & NA    & -0.20 & -0.01 & NA    & NA    & NA    & 0.94  & NA    & NA    & NA    & NA   \\
  \end{mytab}

\explain{Each cell is the slope of a time-series regression of the log number of cyclones on a year index (based on three-year intervals, except for the first period which contains only two years).  When a basin had a zero number in a cell in a three-year period, not slope on log number of cyclones could be calculated.  These cells are marked with ``NA.''}

\end{table}

\begin{table}

  \tblcaption{tbl:statsig}{Statistical Significance Tests}

  \setlength{\tabcolsep}{1.2ex}

  \renewcommand{\t}{\scriptsize}

  \begin{ctabular}{c@{$\;>\;$} c sRRR s RRR}
    \toprule
    \mc2c{}        & \mc3c{3/2 First Period} & \mc3c{Omit First Period}                                \\
    \cmidrule(lr){3-5} \cmidrule(lr){6-8}
    \mc2c{Compare} & Coef                    & T-Stat & P($>|t|$) & Coef     & T-Stat & P($>|t|$)      \\
    \midrule
    C2             & C1                      & 0.72   & 2.07      & 6.3\%    & 0.58   & 1.45  & \t17.9\% \\
    C3             & C1                      & 1.47   & 3.49      & 0.5\%    & 1.08   & 2.57  & 2.8\%  \\
    C4             & C1                      & 1.08   & 2.92      & 1.4\%    & 1.05   & 2.41  & 3.7\%  \\
    C5             & C1                      & -0.42  &  -0.46    & \t65.7\% & -0.16  & -0.15 & \t88.4\% \\
    \midrule
    C3             & C2                      & 0.75   & 2.38      & 3.7\%    & 0.51   & 1.51  & \t16.2\% \\
    C4             & C2                      & 0.36   & 1.67      & \t12.2\% & 0.48   & 1.97  & 7.7\%  \\
    C5             & C2                      & -1.14  & -1.34     & \t20.8\% & -0.74  & -0.76 & \t46.4\% \\
    \midrule
    C4             & C3                      & -0.40  & -1.13     & \t28.4\% & -0.03  & -0.09 & \t92.6\% \\
    C5             & C3                      & -1.90  & -1.98     & 7.3\%    & -1.25  & -1.20 & \t25.9\% \\
    \midrule
    C5             & C4                      & -1.50  & -1.76     & \t10.6\% & -1.21  & -1.23 & \t24.8\% \\
    C5             & C3+C4                   & -1.70  & -1.92     & 8.2\%    & -1.24  & -1.23 & \t24.6\% \\
    \addlinespace
    C3+C4          & C1+C2                   & 1.09   & 3.82      & 0.3\%    & 0.91   & 2.90  & 1.6\%  \\
    C3+C4+C5       & C1+C2                   & 0.95   & 3.48      & 0.5\%    & 9.82   & 2.64  & 2.5\%  \\
    \bottomrule
  \end{ctabular}

  \explain{Higher category cyclones are in the left column, lower category cyclones are in the right column.  The test is whether the slope of the higher-category is greater than the slope of the lower category.  If it is positive, cyclones became relatively more intense.   The Kossin slope difference (increase in log C3-C5 relative to C1-C2 cyclones) is highly statistically significant, with a T-statistic of $+3.48$.  The log number of C5 cyclones slope coefficient relative to the equivalent of C3 and C4 cyclones has a T-statistic of $-1.92$.  Significance in the right column is smaller.  Significance levels above 10\% are in smaller fonts.}

\end{table}

\begin{figure}[h]

  \figcaption{fig:n}{\textcite[Figure 3]{kossin2020global}, Components}

  \fourfig{n}

  \explain{The blue dashed line are the log-number of C1-C2 cyclones (scale on the right).  The red dashed line are the log number of C3-C5 cyclones.}

\end{figure}

\appendix
\begin{table}[h]
  \newenvironment{mysub}{
    \begin{tabular}{csRs *{5}{R}}
      \toprule
           & All   & C1  & C2  & C3  & C4  & C5 \\
      \midrule}{\bottomrule\end{tabular}}

  \tblcaption{tbl:desc-by-year}{APPENDIX: Sample Mean By Period and Percentage Change}

  \small

  \newenvironment{ctwobase}{
    \begin{ctabular}{cssc}
    }{\end{ctabular}}

  \begin{ctwobase}
    First Third (1979-1990) & Last Third (2006-2017) \\[1ex]
    \begin{mysub}
      ALL & 1,464 & 735 & 256 & 229 & 197 & 46 \\
      \midrule
      NA  & 115   & 74  & 17  & 12  & 10  & 1  \\
      SI  & 189   & 117 & 29  & 26  & 18  & 0  \\
      SP  & 189   & 95  & 37  & 32  & 21  & 4  \\
      EP  & 392   & 224 & 61  & 58  & 46  & 4  \\
      WP  & 537   & 196 & 109 & 96  & 100 & 37 \\
      NI  & 42    & 30  & 3   & 6   & 2   & 0  \\
    \end{mysub}
    &
    \begin{mysub}
      ALL & 1,469 & 674 & 272 & 261 & 233 & 29 \\
      \midrule
      NA  & 229  & 99  & 44  & 40  & 41  & 5  \\
      SI  & 231  & 117 & 36  & 47  & 30  & 1  \\
      SP  & 142  & 74  & 22  & 25  & 19  & 2  \\
      EP  & 333  & 164 & 53  & 57  & 50  & 9  \\
      WP  & 504  & 194 & 115 & 91  & 92  & 13 \\
      NI  & 30   & 25  & 2   & 1   & 2   & 0  \\
    \end{mysub} \\[2em]
    \addlinespace
    \\
    Last Third (2006-2017) & Percentage Change 1979-1990 to 2006-2017 \\[1ex]
    \begin{mysub}
      ALL & 1,326 & 572   & 241   & 266   & 212   & 34    \\
      \midrule
      NA  & 198   & 89    & 38    & 42    & 27    & 2     \\
      SI  & 189   & 93    & 29    & 38    & 28    & 1     \\
      SP  & 113   & 49    & 22    & 22    & 17    & 4     \\
      EP  & 336   & 168   & 54    & 58    & 52    & 4     \\
      WP  & 448   & 146   & 94    & 100   & 85    & 23    \\
      NI  & 41    & 27    & 4     & 6     & 3     & 0     \\
    \end{mysub}
    &
    \renewcommand{\tabcolsep}{3pt}
    \begin{mysub}
      ALL & -0.09 & -0.22 & -0.06 & 0.16  & 0.08  & -0.26 \\
      \midrule
      NA  & 0.72  & 0.20  & 1.24  & 2.50  & 1.70  & 1.00  \\
      SI  & 0.00  & -0.21 & 0.00  & 0.46  & 0.56  & Inf   \\
      SP  & -0.40 & -0.48 & -0.41 & -0.31 & -0.19 & 0.00  \\
      EP  & -0.14 & -0.25 & -0.11 & 0.00  & 0.13  & 0.00  \\
      WP  & -0.17 & -0.26 & -0.14 & 0.04  & -0.15 & -0.38 \\
      NI  & -0.02 & -0.10 & 0.33  & 0.00  & 0.50  & -     \\
    \end{mysub}
  \end{ctwobase}

\end{table}

\end{document}